\def\year{2016}\relax
\documentclass[letterpaper]{article}
\usepackage{aaai16}
\usepackage{times}
\usepackage{helvet}
\usepackage{courier}
\usepackage[capposition=top]{floatrow}
\usepackage{multirow}
\usepackage{caption}
\usepackage{amsmath}
\usepackage{graphicx}
\newcommand*{\TitleFont}{%
      \usefont{\encodingdefault}{\rmdefault}{b}{n}%
      \fontsize{13.97}{100}%
      \selectfont}
\title{\TitleFont Catching Fire via `Likes': Inferring Topic Preferences of Trump Followers on Twitter}

\nocopyright{}
 
\begin{document}

\author{Yu Wang\\Political Science\\University of Rochester\\Rochester, NY 14627\\ywang@ur.rochester.edu\And Jiebo Luo\\Computer Science\\University of Rochester\\Rochester, NY 14627\\jluo@cs.rochester.edu\And Richard Niemi\\Political Science\\University of Rochester\\Rochester, NY 14627\\niemi@rochester.edu \And Yuncheng Li\\Computer Science\\University of Rochester\\Rochester, NY 14627\\yli@cs.rochester.edu \And Tianran Hu\\Computer Science\\University of Rochester\\Rochester, NY 14627\\thu@cs.rochester.edu}

\maketitle

\begin{abstract} \small\baselineskip=9pt 
In this paper, we propose a framework to infer the topic preferences of Donald Trump's followers on Twitter. We first use latent Dirichlet allocation (LDA) to derive the weighted mixture of topics for each Trump tweet. Then we use negative binomial regression to model the ``likes,'' with the weights of each topic serving as explanatory variables. Our study shows that attacking Democrats such as President Obama and former Secretary of State Hillary Clinton earns Trump the most ``likes.'' Our framework of inference is generalizable to the study of other politicians.

\end{abstract}

\section{Introduction} 
Republican presidential front-runner Donald Trump has 5.46 million followers on Twitter. Between September 18th and December 27nd, 2015, Trump posted an average of 21.0 tweets per day and each Trump tweet received an average of 3410.7 ``likes'' with a standard deviation of 2732.7. In this paper, we analyze these variations in ``likes."

\begin{figure}[H]
\caption{Examples of Trump's Tweets and the ``likes''}
\label{gallery}
\includegraphics[height=7.96cm]{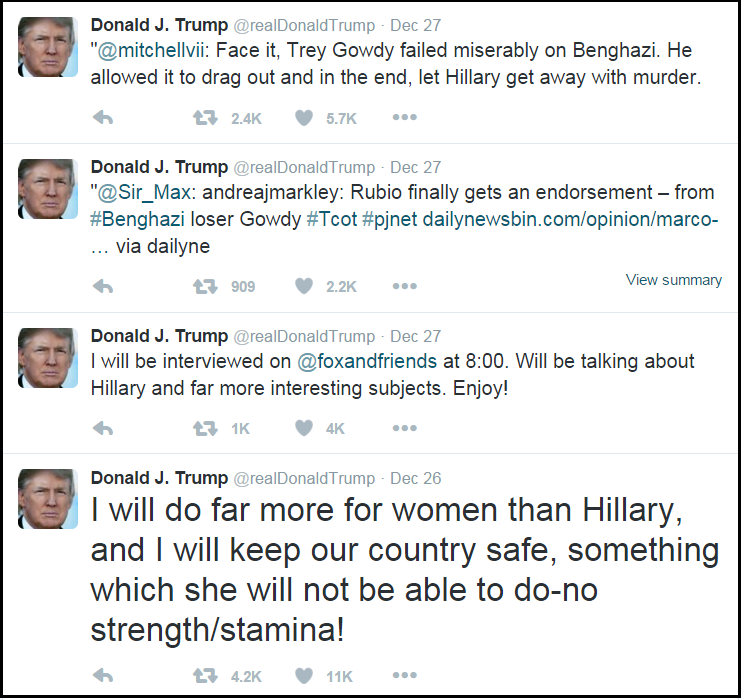}
\end{figure}

Figure \ref{gallery} shows four tweets posted by Donald Trump on December 27, 2015. The first and the fourth of the tweets are about Hillary Clinton, former Secretary of State, who is currently leading in the Democratic presidential race. The second tweet is on Marco Rubio, a presidential candidate from the Republican party.\footnote{For a detailed summary of the many poll results, please see http://elections.huffingtonpost.com/pollster/2016-national-gop-primary.} The third tweet is about Donald Trump appearing in the news media. What motivates our study is the large variations in the number of ``likes.'' In this example, they range from 2,200 to 11,000.

We interpret these ``likes'' as measures of the tweets' attractiveness to Trump's followers: if followers like a Trump tweet, they `like' it; otherwise, they do not. We further assume that each tweet as a document represents a mixture of topics and that the variations in these ``likes'' can be partly attributed to these topics, thus revealing the followers' preference. The most preferred topic is expected to earn Trump the most ``likes.''
 
To operationalize this idea, we first apply latent Dirichlet allocation (LDA) to extract the topics \cite{lda,LDA2}. Then we use negative binomial regression to model the ``likes,'' with the weights of each topic serving as explanatory variables. Throughout, we shall control for other factors that influence the number of ``likes'': time of the posting, weekend, public debates and the number of Trump followers.

%We design our inference procedure as follows. First apply latent dirichlet analysis (LDA)to extract the topics.
The contributions of our study are threefold. First we contribute to the study of social behavior on Twitter. The existing literature has mostly focused on analyzing the retweeting behavior. Our study complements this literature by exploring the ``like'' behavior. Second, our study shows that the most favored topic for the Trump followers is attacking the Democrats. This result shall prove valuable for both political scientists and politicians. Third, we propose a framework for inferring followers' topic preference. This framework can be applied to the study of other politicians.
 
\section{Related Work}

Our work builds upon previous research on electoral studies using social media data and on behavioral studies in social media.

There are a large number of studies on using social media data to analyze and forecast election results. \cite{moretweetsmorevotes} finds a statistically significant relationship between tweets and electoral outcomes. \cite{facebookCongress} suggests that a candidate's number of ``likes'' in Facebook can be used for measuring a campaign's success in engaging the public. According to \cite{fbcount}, the number of Facebook fans constitutes an indicator of candidate viability. \cite{trumpists} uses the profile images on Twitter to study the demographic characteristics of Donald Trump's and Hillary Clinton's followers. Our work uses both the number of Trump followers (as a control variable) and the number of ``likes'' (as the dependent variable). Our contribution is to infer follower preferences from these ``likes.''

There are also quite a few studies modeling individual behaviors in social media. \cite{retweet} models the decision to retweet, using Twitter user features such as agreeableness, number of tweets posted, and daily tweeting patterns. \cite{answer} models individuals' waiting time before replying to a tweet based on their previous replying patterns. Our study models the number of ``likes'' that a Trump's tweet receives. Our innovation is to use tweet-specific features instead of individual-specific features, as done in the above-cited literature.
%Research by \cite{tumasjan} finds that the number of messages mentioning a party reflects the election results.
\section{Data and Methodology}
\begin{figure}[H]
\caption{(Time series) Number of Trump Followers}
\label{followers}
\includegraphics[height=4cm]{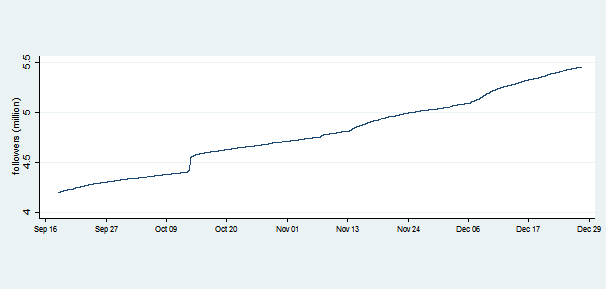}
\end{figure}

\begin{figure}[H]
\caption{(Time series) `Likes' per Trump Tweet}
\label{tweetLikes}
\includegraphics[height=3.6cm]{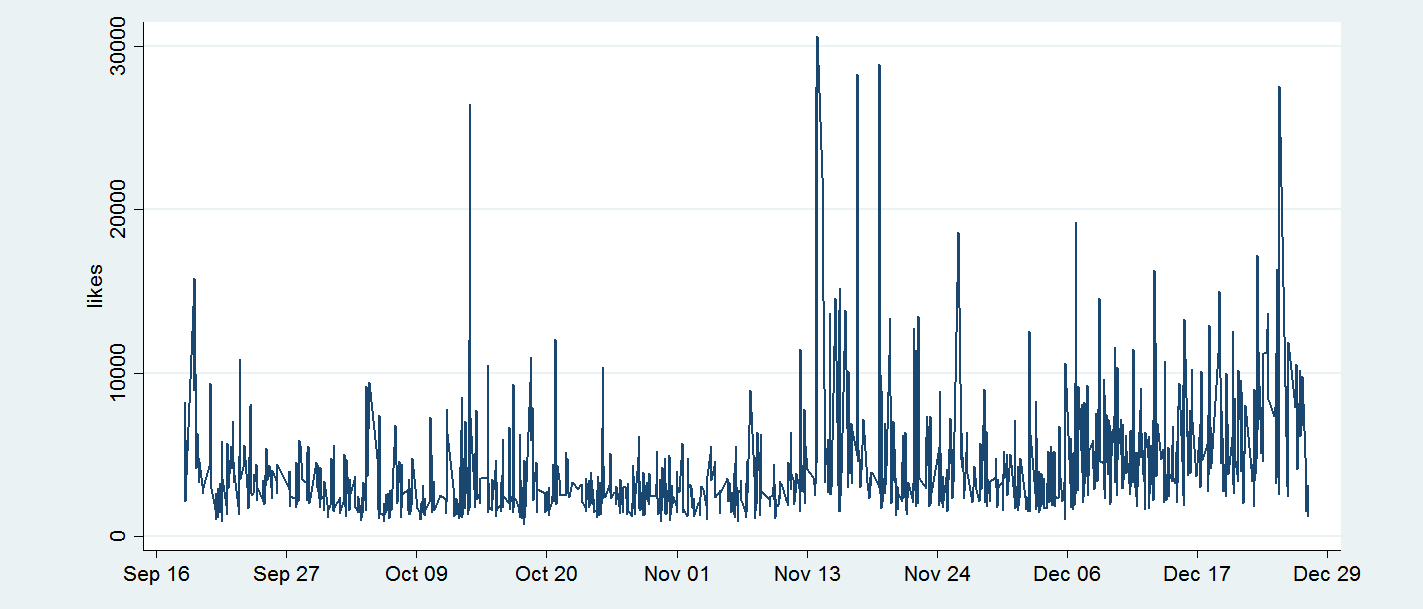}
\end{figure}

We use the dataset $\textit{US2016}$, constructed by us with Twitter data. The dataset contains a tracking record of the number of followers for all the major candidates in the 2016 presidential race, including Donald Trump (Figure \ref{followers}). The dataset spans the entire period between September 18th, 2015 and December 27th, 2015 and covers three Democratic debates and three Republican debates. In addition, $\textit{US2016}$ also contains all the tweets (2120, in total) that Trump posted during the same period and the number of ``likes'' that each tweet has received (Figure \ref{tweetLikes}). In Table \ref{sumstat}, we report the summary statistics of the main variables.

\begin{table}[htbp]\centering \caption{Summary statistics \label{sumstat}}
\setlength{\tabcolsep}{3pt}
\begin{tabular}{l c c c c  c}\hline\hline
\multicolumn{1}{c}{\textbf{Variable}} &\textbf{Min}& \textbf{Max}& \textbf{Mean}
&\textbf{S.D.} & \textbf{N}\\ \hline
Likes & 741& 30612&  3411 & 2733  & 2120\\
Democratic Debates & 0 & 1& 0.088 & 0.283  & 2120\\
Republican Debates &0 & 1& 0.072 & 0.258  & 2120\\
Followers (million) &4.50& 5.45& 4.788 & 0.344  & 2064\\
\hline
\end{tabular}
\end{table}

LDA and negative binomial regression are the two workhorses of our work. We first use LDA to extract topics from the tweets and then use negative binomial to estimate the coefficients for each topic. In our estimation, we use the following two link functions:

\begin{equation*} \label{eq1}
\begin{split}
\mu &=exp(\beta_0+ \beta_1\mathrm{Weekend}+\beta_2\mathrm{Democratic\: Debates}\\
 &+\beta_3\mathrm{Republican\: Debates}+\beta_4\mathrm{Follower\:Count}\\
 &+\pmb{\gamma}\cdot \mathbf{Topic}+\theta\cdot\mathbf{Hour\:Controls})\\
 p& =1/(1+\alpha\mu)\\
\end{split}
\end{equation*}

where \textit{Democratic Debates} is binary and takes the value of 1 on the day of a Democratic debate and on the day immediately after the debate, \textit{Republican Debates} is binary and takes the value of 1 on the day of a Republican debate and on the following day. \textit{Weekend} is binary and is 1 if the tweet is posted during the weekend. \textit{Follower Count} is the number of Trump followers when the tweet is posted. \textbf{Topic} is a vector and denotes the weights on each topic. \textbf{Hour Controls} are a set of dummy variables controlling for the hours when the tweet is posted. $\alpha$ controls the gamma distribution, Gamma(1/$\alpha$, $\alpha$), that generates over-dispersion.

\iffalse
\begin{equation*}
\begin{split}
p=1/(1+\alpha\mu)
\end{split}
\end{equation*}
\fi
With these two equations, we formulate the likelihood as follows and use it to estimate the coefficients:
\[\frac{\Gamma(1/\alpha+y)}{\Gamma(y+1)\Gamma(1/\alpha)}p^{1/\alpha}(1-p)^y\]
where y denotes the number of ``likes.''

To select the appropriate number of topics for the LDA procedure, we use as a metric the resulting mean absolute error (MAE) of the negative binomial regression.

\begin{figure}[H]
\caption{Mean Absolute Error as a Function of Topics}
\label{mae}
\includegraphics[height=5cm]{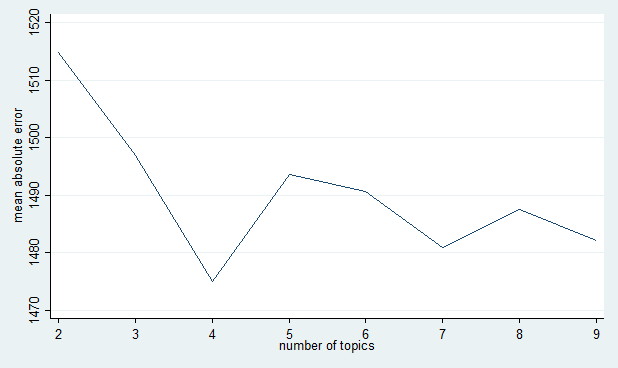}
\end{figure}

In Figure \ref{mae}, we report the MAE as a function of the number of topics, from 2 to 9. We observe that MAE first decreases when the number of topics increases from 2 to 4 and then remains relatively flat. MAE achieves its minimum when the number of topics equals 4, so we shall set the number of topics to 4 in the subsequent analysis.\footnote{Selecting a much larger number of topics, for example 17, yields an even smaller MAE. But such a larger number of topics will make interpretation very difficult.}

\section{Main Results}
In this section, we first present the four topics that we uncover from the LDA procedure. We then report our main estimation results. Lastly, we discuss the estimated differences between these topics.

\subsection{Topic Classification}
In Table \ref{topics}, we report the four topics that we obtain after applying the LDA procedure to the Trump tweets, together with the top 20 topic words. 

We observe these tweets can be classified into four topics: \textit{News Media}, \textit{Republicans}, \textit{Democrats}, and \textit{Trump Campaign}. \textit{News Media} contains such topic words as foxnews, cnn, interview and abc. \textit{Republicans} includes names of Republican candidates such as Jeb Bush, Marco Rubio, Ben Carson and Ted Cruz. \textit{Democrats} includes names from the Democratic party, such as Hillary Clinton and President Obama. \textit{Trump Campaign} relates to Trump's performance in polls and speeches. It includes such words as polls and crowd.

Based on this classification, the first, third, and fourth tweets in Figure \ref{gallery} shall assign large weights to \textit{Democrats} as they all relate to Hillary Clinton. The second tweet assigns a large weight to the topic \textit{Republicans} as it relates to Marco Rubio, a presidential candidate from the Republican party. And the third tweet assigns a large weight  also to the \textit{News Media} topic, as it is about Donald Trump appearing in the news. 

\subsection{Estimation}
In this subsection, we use the weights derived from the LDA procedure to estimate the topic preferences of the Trump followers on Twitter. We follow the formulation presented in Section 3 and we report the estimation results in Table \ref{nb}.

Column 1 does not include the extracted topics as explanatory variables and serves as the baseline. It shows that tweets Trump posted during Democratic debates tend to receive more ``likes'' and tweets posted during Republican debates tend to receive fewer ``likes.'' A similar observation is reported in \cite{trumpists}. The coefficient on \textit{Follower Count} is positive and statistically significant, suggesting that the more followers Trump has the more ``likes'' a Trump tweet will receive.  We also find that tweets posted during the weekend tend to receive fewer ``likes.''

By contrast, Column 2 includes the topic variables. First, by the Akaike Information Criterion (AIC), we shall select the Column 2 specifications. Second, we exclude the topic \textit{News Media} from the regression to avoid perfect multi-collinearity as the four topics sum up to 1. We find that the topics \textit{Democrats} and \textit{Trump Campaign} are statistically more preferred to \textit{News Media}. The difference between \textit{Republicans} and \textit{News Media} is not statistically significant.

\begin{table}[H]\centering
\setlength{\tabcolsep}{13.3pt}
\renewcommand{\arraystretch}{1.1}
\def\sym#1{\ifmmode^{#1}\else\(^{#1}\)\fi}
\caption{Negative Binomial Regression}
\label{nb}
\begin{tabular}{l*{2}{c}}
\hline\hline
                    &\multicolumn{1}{c}{Baseline}&\multicolumn{1}{c}{Topics}\\
\hline
likes               &                     &                     \\
Democratic Debates            &       0.365\sym{***}&       0.286\sym{***}\\
                    &    (0.0423)         &    (0.0412)         \\
Republican Debates            &      -0.225\sym{***}&      -0.165\sym{***}\\
                    &    (0.0479)         &    (0.0463)         \\
Follower Count    & 0.691\sym{***}&       0.629\sym{***}\\
                    &  (0.0335)         &    (0.0326)         \\
Weekend             &      -0.135\sym{***}&      -0.113\sym{***}\\
                    &    (0.0269)         &    (0.0260)         \\
Republicans         &                     &      0.0368         \\
                    &                     &    (0.0447)         \\
Democrats           &                     &       0.545\sym{***}\\
                    &                     &    (0.0456)         \\
Trump Campaign            &                     &      0.0840\sym{*}  \\
                    &                     &    (0.0421)         \\
Constant            &       4.688\sym{***}&       4.878\sym{***}\\
                    &     (0.217)         &     (0.212)         \\
\hline
ln($\alpha$)             &                     &                     \\
Constant            &      -1.270\sym{***}&      -1.346\sym{***}\\
                    &    (0.0298)         &    (0.0299)         \\
\hline
Observations        &        2063         &        2063         \\
\textit{AIC}        &     36312.5         &     36148.2         \\
\hline\hline
\multicolumn{3}{l}{\footnotesize Standard errors in parentheses}\\
\multicolumn{3}{l}{\footnotesize \sym{*} \(p<0.05\), \sym{**} \(p<0.01\), \sym{***} \(p<0.001\)}\\
\end{tabular}
\end{table}

Using likelihood ratio test on $\alpha$, we are further able to confirm the existence of over-dispersion, and thus confirm that negative binomial regression is more appropriate than Poisson regression. 

\begin{table*}[]
\renewcommand{\arraystretch}{1}
\setlength{\tabcolsep}{3.5pt}
\centering
\caption{Topic Classification}
\label{topics}
\begin{tabular}{ll}
\hline\hline
Topics                        & Top 20 Topic Words                                                                            \\
\hline
\multirow{2}{*}{News Media}   & trump great america thank again donald make foxnews makeamericagreatagain                      nice thanks cnn interview\\
                              & president good more tonight job vote need \\\hline %abc                      
                               
\multirow{2}{*}{Republicans}  & trump cnn poll rubio carson won jeb debate bush jebbush campaign gop marco immigration  people \\
                              & foxnews still candidate megynkelly money \\\hline%polls like cruz even establishment vote              
\multirow{2}{*}{Democrats}    & hillary president why people obama like many going get isis right clinton country want       need doing them\\
                              & against foxnews last \\\hline %last think never how more use better                         
\multirow{2}{*}{Trump Campaign} & trump great new poll thank makeamericagreatagain live donald crowd america iowa               big night morning \\
                              & book cnn tonight people tomorrow polls \\\hline\hline %today                               
                        
\end{tabular}
\end{table*}

\subsection{The Most Preferred Trump Topic}
To visualize the results, we plot the estimated coefficients with 95\% confidence intervals in Figure \ref{topics}. \textit{News Media}, taking the value of 0, serves as the baseline for comparison. The confidence intervals for \textit{Democrats} and \textit{Trump Campaign} are both above 0, suggesting that Trump followers prefer \textit{Democrats} and  \textit{Campaign} to \textit{News Media}. By comparison, the confidence interval for \textit{Republicans} is not strictly positive. 

More important, we observe that \textit{Democrats} is the most preferred topic among Trump followers, which is consistent with our earlier observation that tweets posted during Democratic debates tend to receive more ``likes.'' Referring back to the quoted tweets in Figure \ref{gallery}, this helps explain why tweets mentioning Hillary Clinton receive more ``likes.''
% ``News media'' is about Trump appearing in the media. Tweets such as ``I will be interviewed on the @TODAYshow at 7:00 A.M. Talking about the state of the race for president, and more. Enjoy!'' belong to this catetory. 

% \usepackage{multirow}

\begin{figure}[H]
\caption{Estimated Topic Coefficients (95\% C.I.)}
\label{dailyPattern}
\includegraphics[height=6.5cm,width=8.4cm]{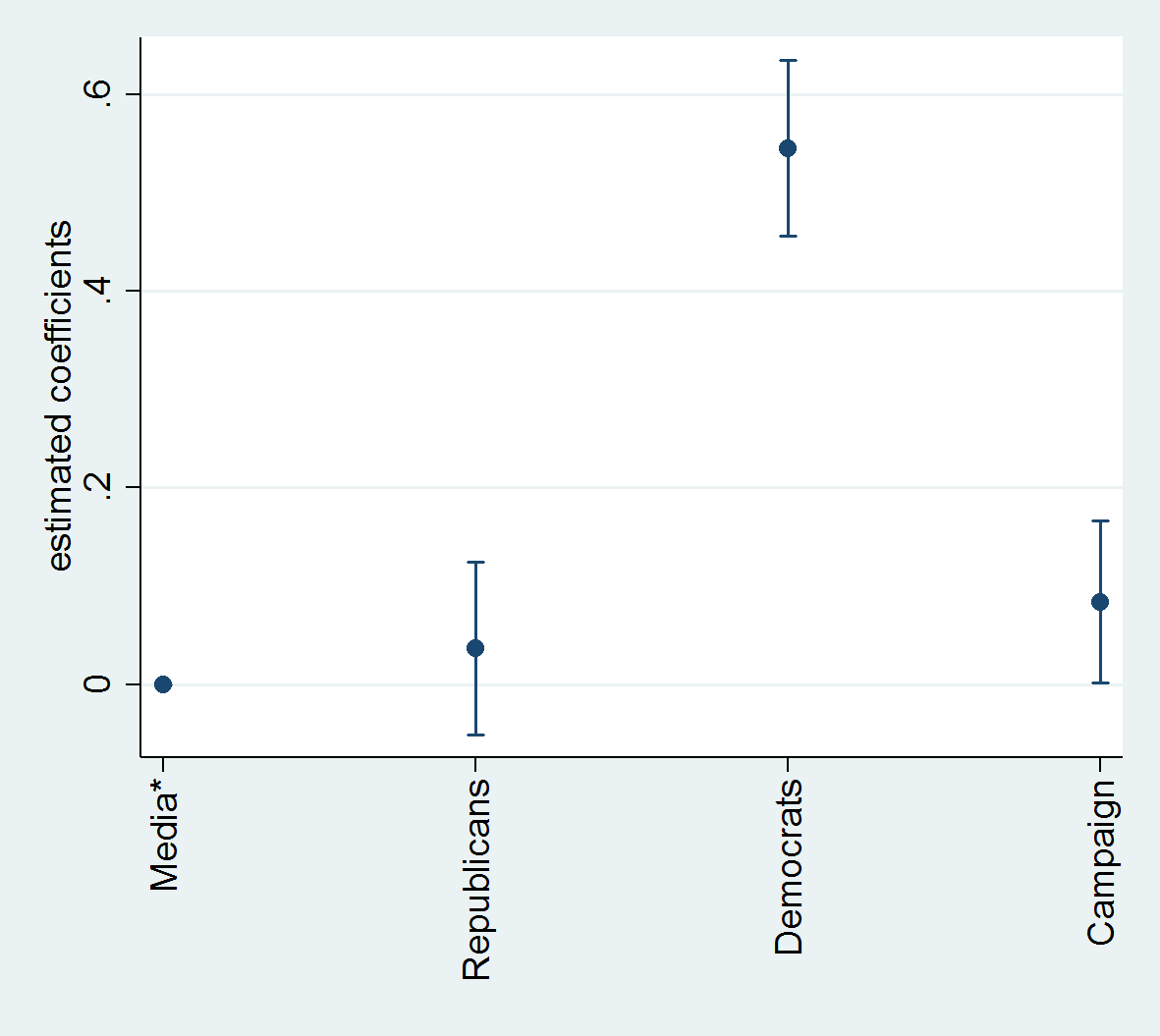}
\end{figure}

\section{Conclusion and Future Work}
We have presented a study of the topic preferences of the presidential candidate Donald Trump's followers on Twitter. We first used latent Dirichlet allocation (LDA) to extract topics from Trump's tweets and then regressed the number of ``likes'' on the weighted mixture of topics. We used Mean Absolute Error (MAE) from the negative binomial regression as a metric to select the appropriate number of topics. We found that \textit{Democrats} is the most preferred topic among Trump followers. While our work focuses on Donald Trump, the framework of inference that we propose here can be applied to the study of other politicians, such as Hillary Clinton and President Barack Obama.

We believe the rise of Donald Trump is a significant event in American politics. Our immediate next step is to understand the demographics of the Trump followers on Twitter and evaluate their sentiments through, e.g., tweets.

\section{Acknowledgment}
We gratefully acknowledge support from the University of Rochester, New York State through the Goergen Institute for Data Science, and our corporate sponsors Xerox and Yahoo. 

\bibliographystyle{aaai}
\bibliography{DREAM-TRIO}
\end{document}